\documentclass[singlespacing]{elsart}
\usepackage{amssymb,graphicx}
\journal{Physica A}
\begin{document}
\begin{frontmatter}

\title{Testing a priority-based queue model with Linux command histories}
\author{Seung Ki Baek},
\author{Tae Young Kim},
\author{Beom Jun Kim\corauthref{cor}}
\corauth[cor]{Corresponding author.}
\ead{beomjun@skku.edu}
\address{Department of Physics, BK21 Physics Research Division, and
Institute of Basic Science, Sungkyunkwan University, Suwon 440-746,
Korea}

\begin{abstract}
We study human dynamics by analyzing Linux history files. The goodness-of-fit
test shows that most of the collected datasets belong to the universality class
suggested in the literature by a variable-length queueing process based on
priority. In order to check the validity of this model, we design two tests
based on mutual information between time intervals and a mathematical
relationship known as the arcsine law. Since the previously suggested queueing
process fails to pass these tests, the result suggests that the modelling of
human dynamics should properly consider the statistical dependency in the
temporal dimension.
\end{abstract}
\begin{keyword}
Waiting-time distribution \sep Power-law
\sep Linux \sep Human dynamics
\PACS 89.75.Da \sep 05.45.Tp \sep 02.50.Ey
\end{keyword}
\end{frontmatter}

\section{Introduction}
Recently, there have been various attempts to characterize the human behaviors
in mathematical terms which have been successfully applied to natural
phenomena.
To name a few, the human activities like Internet,
traffic flows, family names and stock prices are under active
investigation and give deep insights into our
society~\cite{albert,kujawski,wang,bjkim,baek,kiet}.
Now we have even a popularized term known as `human dynamics'~\cite{barabasi},
and many researchers are devoting themselves to this field.
One of their most surprising claims is that there exist
a few universality classes in human
dynamics~\cite{oliveira,vazquez2005,vazquez2006}.
Those universality classes are described by a priority-based queuing
process, which yields power-law waiting-time distributions $p(\tau)\sim
\tau^{-\alpha}$ with universal exponents of $\alpha = 1.0, 1.5$, and
$2.5$~\cite{grinstein}.
This idea has generated great public attention due to its philosophical
implications against our conventional belief in human conditions,
and there have been also intensive scientific debates on their
observations~\cite{stouffer-barabasi-r} and on the existence of
universality classes~\cite{tao}.
Even if one may doubt the validity
of the universality claim, their original observation truly pointed out some
fundamental properties of human behaviors and contributed a lot to this
field by proposing a powerful and falsifiable model.
To our knowledge, only a few models are yet to undergo closer examinations,
including those in Refs.~\cite{Vazquez2006b,blanchard}.

In this article, we analyze human behaviors through Linux history files,
which contain the histories of every shell command input by terminals.
Unlike the records in supercomputers~\cite{kleban} and personal computers
including mouse movements~\cite{slijper},
our observation partially supports the universality claim in that most of
the collected distributions fall into the suggested universality class
with $\alpha=1.5$.
Since this is the regime where a priority-based queue model
works with varying the queue length~\cite{oliveira},
we may imagine that a person works as the model
describes, where each command executed on the shell introduces the next
command to her queue with a randomly assigned priority.
The waiting time before execution is essentially dominated by
a random walk of the queue length,
which gives the desired power-law distribution with
$\alpha=1.5$~\cite{newman}.
However, the overall distribution shows only a small amount of
information and it is more than possible to devise further
examinations to compare our empirical data with the suggested model.
In other words, if command inputs can be
described by the queue model which reduces to a one-dimensional
random walker, such a simple and rigorous mechanism
should put some explicitly testable constraints on the result.
For example, a natural requirement is that the time intervals
between two consecutive events must be mutually independent of each other.
That is our motivation to design two tests based on the correlation between
time intervals and the characteristic hitting time distribution
as a regenerative process~\cite{ross}, respectively. These tests prove that
our observations are not fully explained by the existing queue models.

This article is organized as follows: In Sec.~\ref{sec:collect}, we explain
how we prepared datasets and present their basic statistical features.
In Sec.~\ref{sec:analysis}, the goodness-of-fit test for verifying power-law
behaviors is followed by two tests to examine the priority-based queue model.
Then we discuss the implications of the test results in
Sec.~\ref{sec:discuss} and conclude this work in Sec.~\ref{sec:conc}.

\section{Data Collection}
\label{sec:collect}
\begin{figure}
\includegraphics{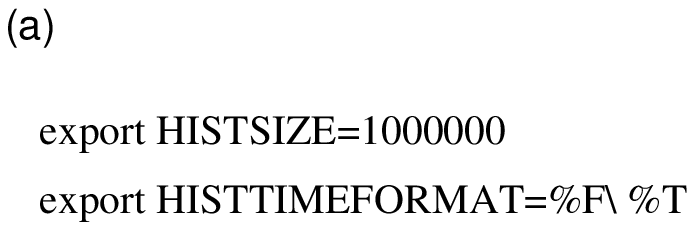}\\
\includegraphics{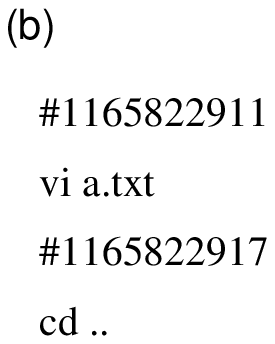}
\caption{(a) Two lines added to~.bashrc to record time-stamps up to
$10^6$ lines. (b) A typical look at the resulting history file, which
contains one time-stamp above every command input to the Linux shell.}
\label{fig:file}
\end{figure}

\begin{figure}
\includegraphics[width=.8\textwidth]{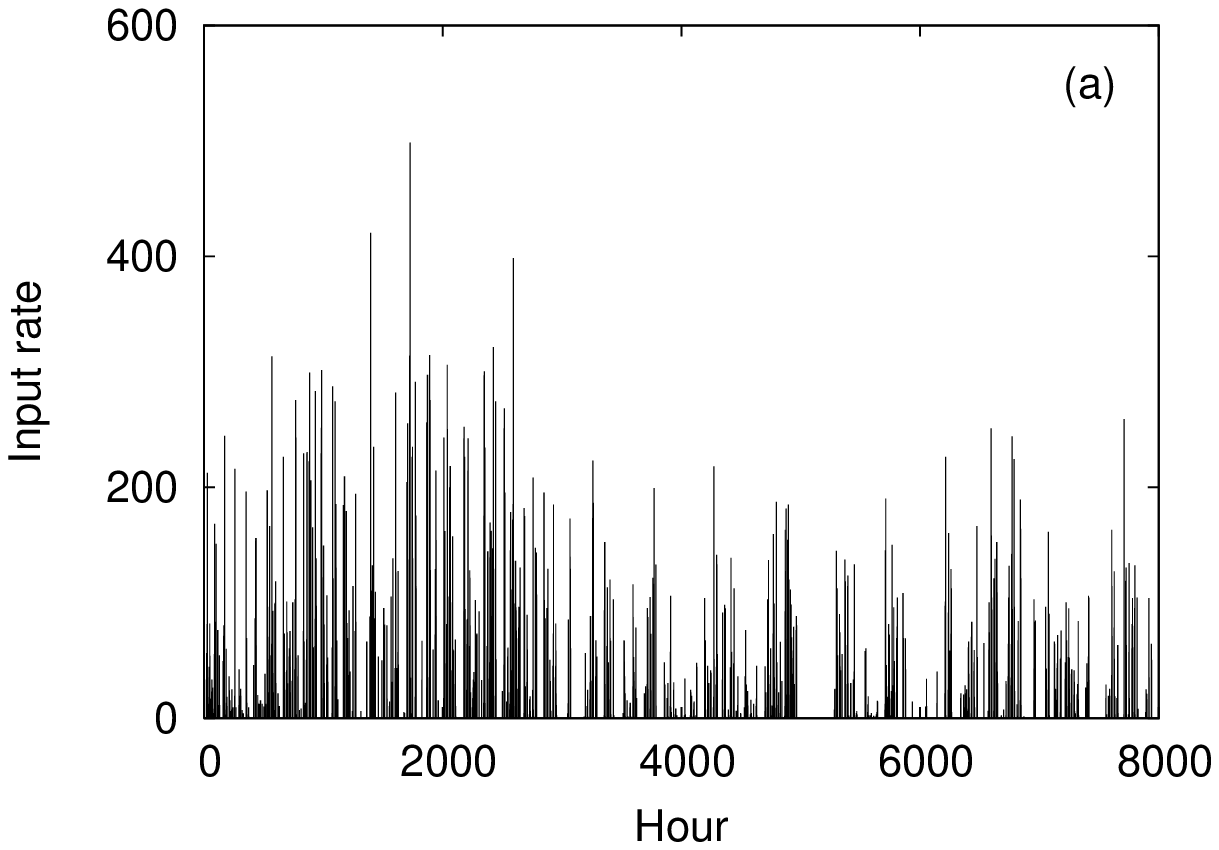}
\includegraphics[width=.8\textwidth]{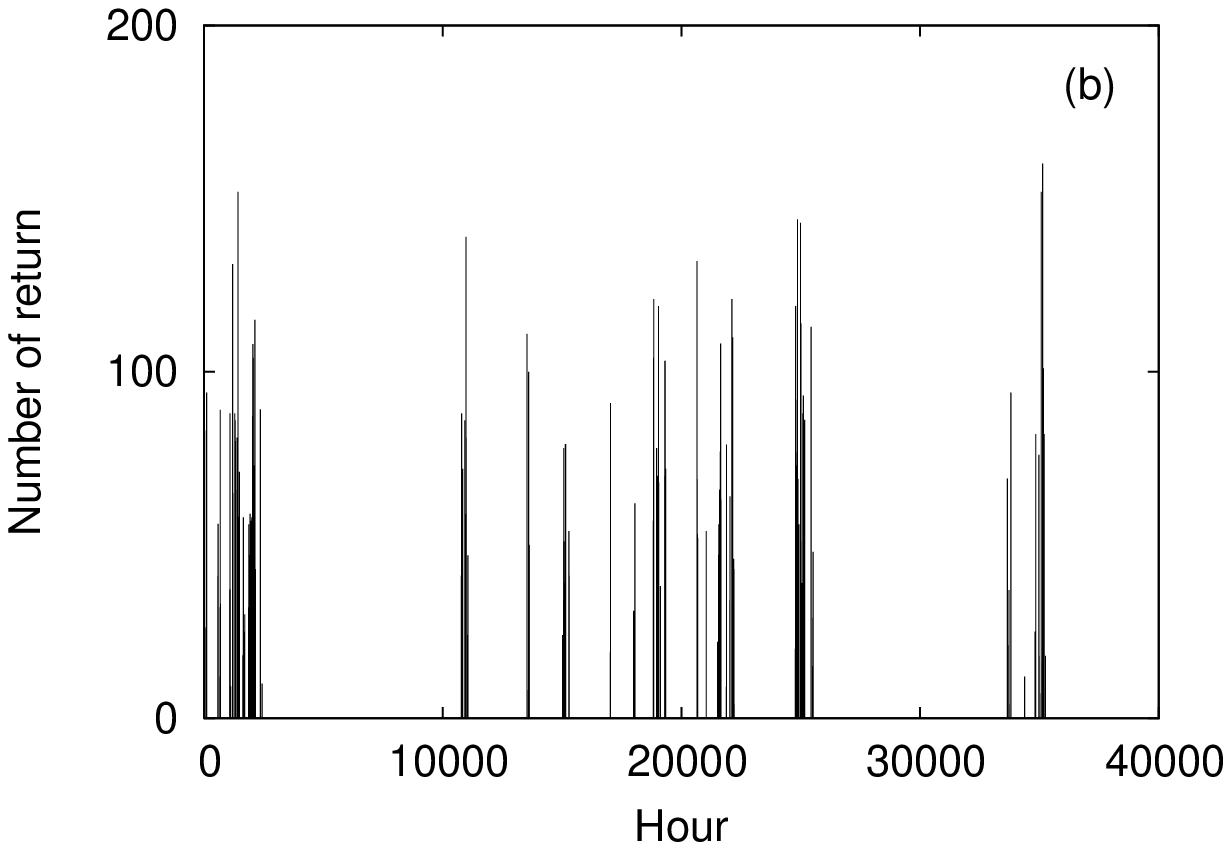}
\caption{(a) Command input rate of the dataset $A$, measured by the number
of inputs in every hour ($=3.6\times10^3$ s).
(b) A one-dimensional random walker's number of return to the origin
in every hour ($=3.6\times10^3$ time steps), from the dataset $R$.}
\label{fig:burst}
\end{figure}

A Linux system usually keeps every user's shell command history up to
some predefined length. In Bash (Bourne-again shell), each shell command can
be made accompanied by a time-stamp,
if we add a couple of lines to a resource file called `.bashrc' as in
Fig.~\ref{fig:file}(a),
where the first line defines the maximal history length and the second adds
time-stamps. Then a user's typical history is written as in
Fig.~\ref{fig:file}(b)
with the numbers indicating time-stamps in units of second.
We collect  seven history files from six users (including
two authors of the present paper),
each of which is given an alphabet from $A$ to $G$.
Since they worked without any explicit coordination during the recording
period, we regard these records as mainly reflecting their individual
characteristics.
Note that the history files may not be arranged in a chronological order
when a user uses multiple terminals so that we have to sort the datasets
before analyses.
In addition, we generate one more dataset $R$, recording the
return times of a one-dimensional random walker to the origin, which
will function as a control group throughout our analyses.
Fig.~\ref{fig:burst} shows the input rates for the
datasets $A$ and $R$ by counting the number of
events in every hour, or in $60^2 = 3600$ time units (seconds).

Letting $t_i$ indicate the time when the $i$th command was entered,
we define the waiting time $\tau_i$ between two consecutive inputs as
\[ \tau_i \equiv t_{i+1} - t_i. \]
Conversely, if there are $n$ command inputs recorded in the file, $T \equiv
\sum_{i=1}^n \tau_i$ is the total time interval. There are two other
characteristic quantities, $\bar\tau \equiv T/n$ and $\tau_{\rm max} \equiv
\max_i \{\tau_i\}$, although we will see later that $\bar\tau$ is not a
good statistic in that it is essentially sample-dependent here.
Those values for each dataset are listed in Table~\ref{table:basic}.

\begin{figure}
\includegraphics[width=\textwidth]{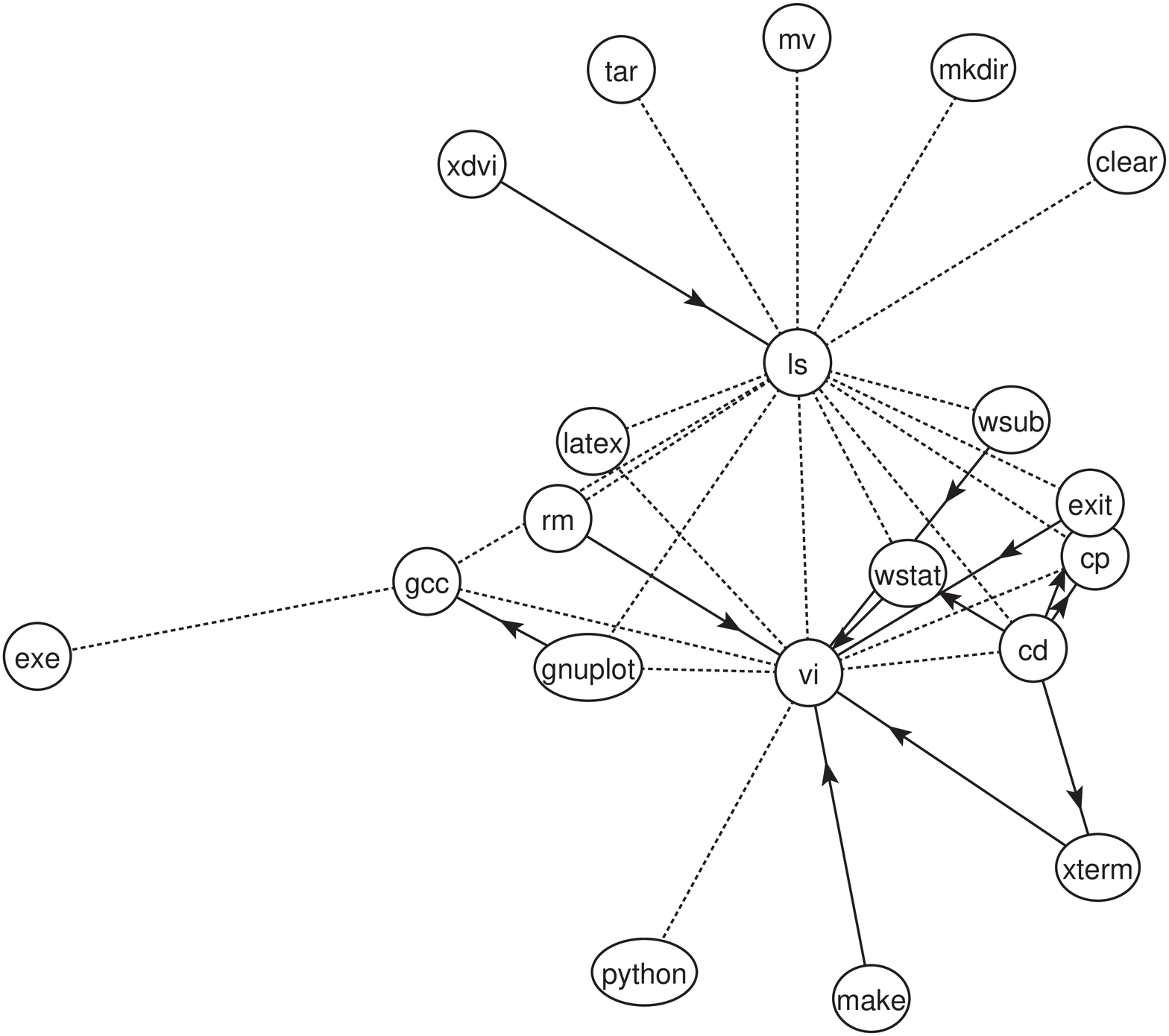}
\caption{Command input pattern of the dataset $A$.
Self-loops are omitted, and the dahsed edges without arrows 
represent the transitions in both directions.
The commands named as `wsub' and `wstat' are not supported by
ordinary Linux shells but specific to this Linux machine,
while `exe' indicates all the user-generated executable files.}
\label{fig:transit}
\end{figure}

Furthermore, one may be interested in the transition between commands.
Suppose that when a command $c_1$ is followed by a command $c_2$
in the shell, we call it a transition from $c_1$ to $c_2$.
The transition patterns are easily visualized by a network, as
shown in Fig.~\ref{fig:transit}  by nodes (commands) and links (transitions).
As the largest frequency is found in the transition from `ls' to `cd',
only those links are depicted whose transition frequencies exceed
$2\%$ of it, together with the major commands involved in these
transitions.

\section{Data Analysis}
\label{sec:analysis}
\subsection{Waiting-time distribution}

\begin{figure}
\includegraphics[width=.5\textwidth]{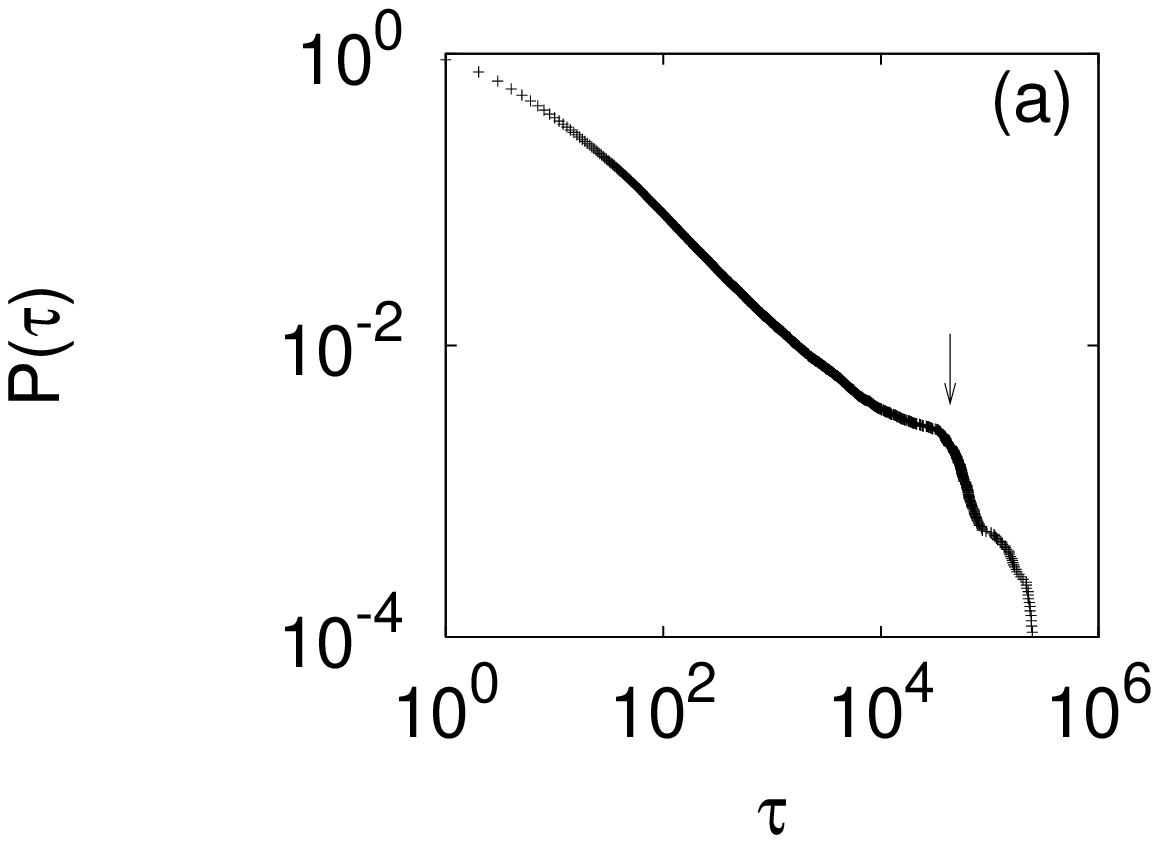}
\includegraphics[width=.5\textwidth]{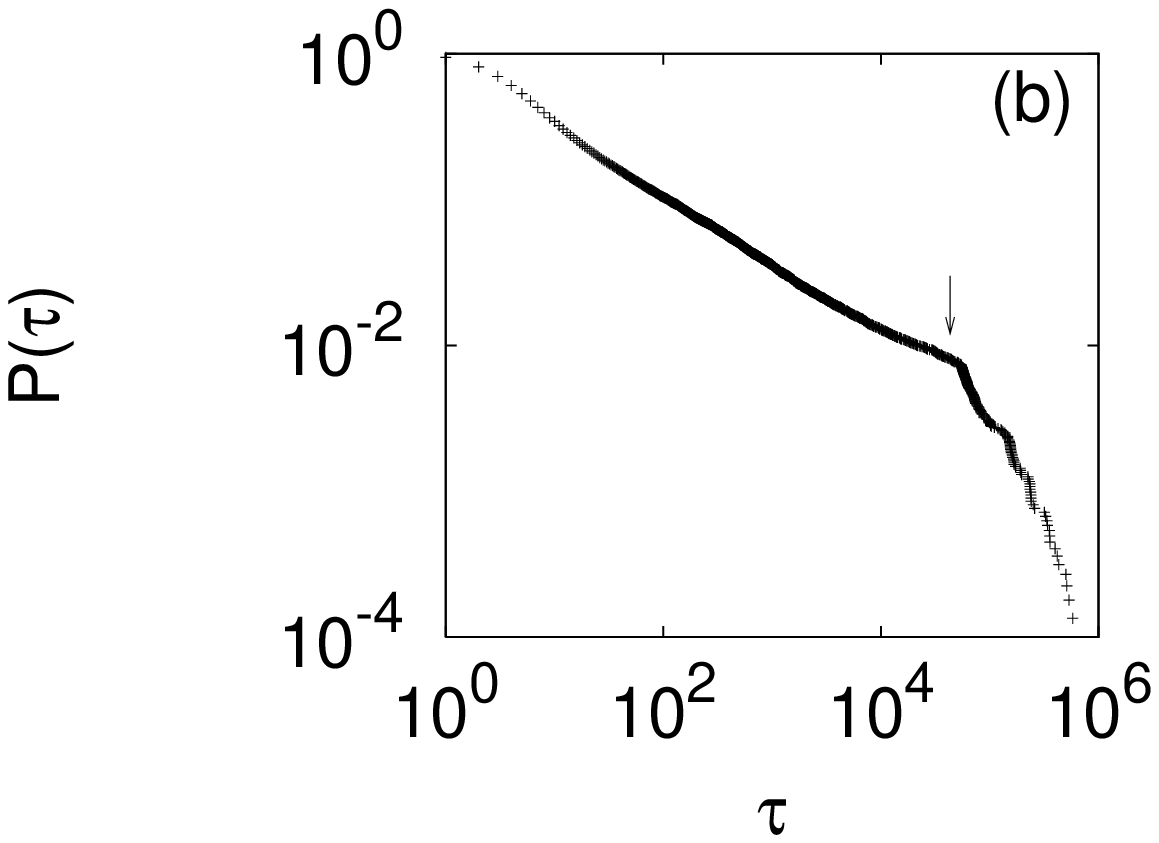}
\includegraphics[width=.5\textwidth]{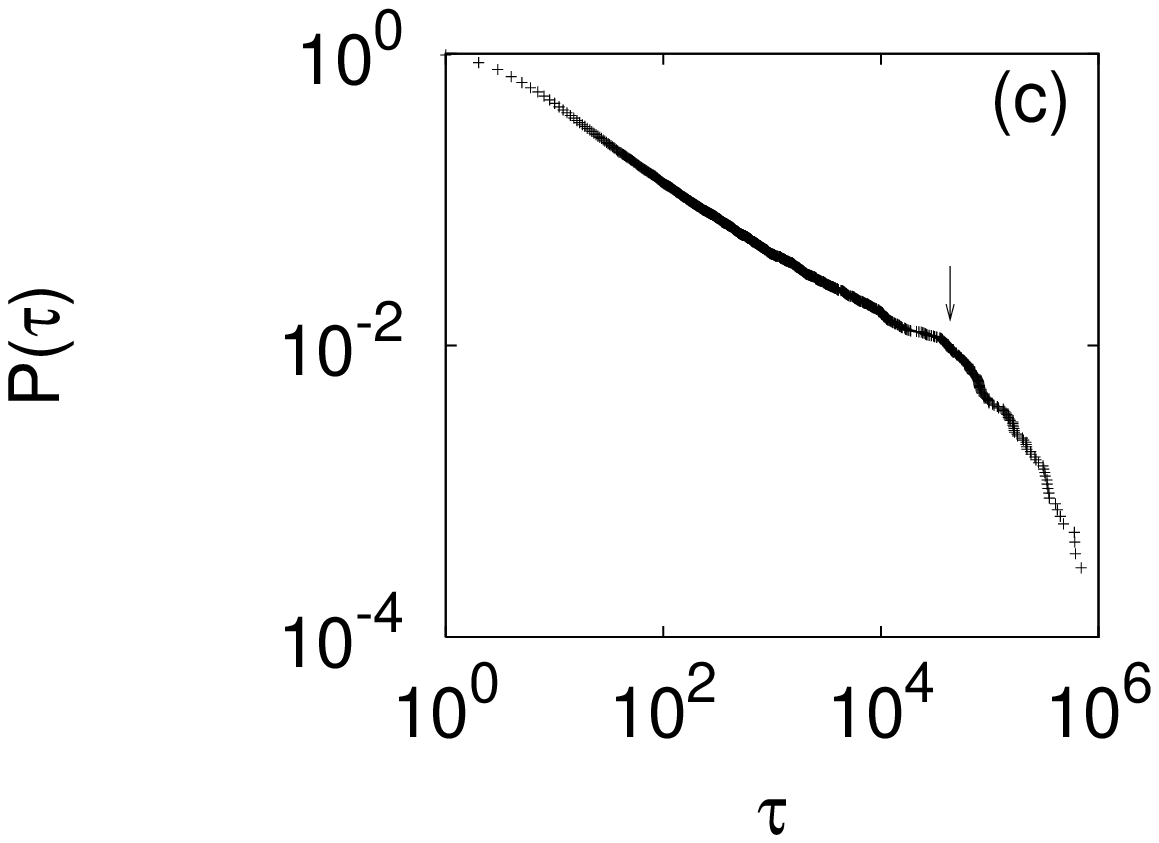}
\includegraphics[width=.5\textwidth]{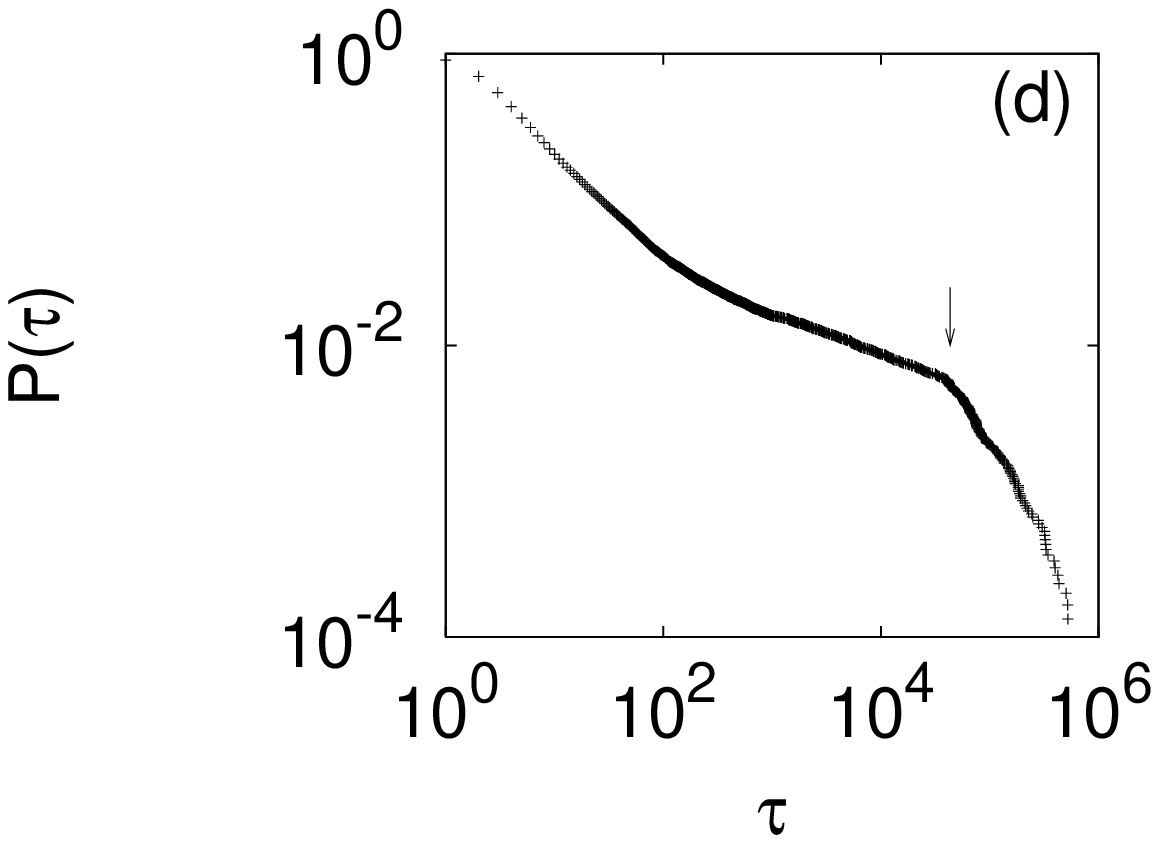}
\includegraphics[width=.5\textwidth]{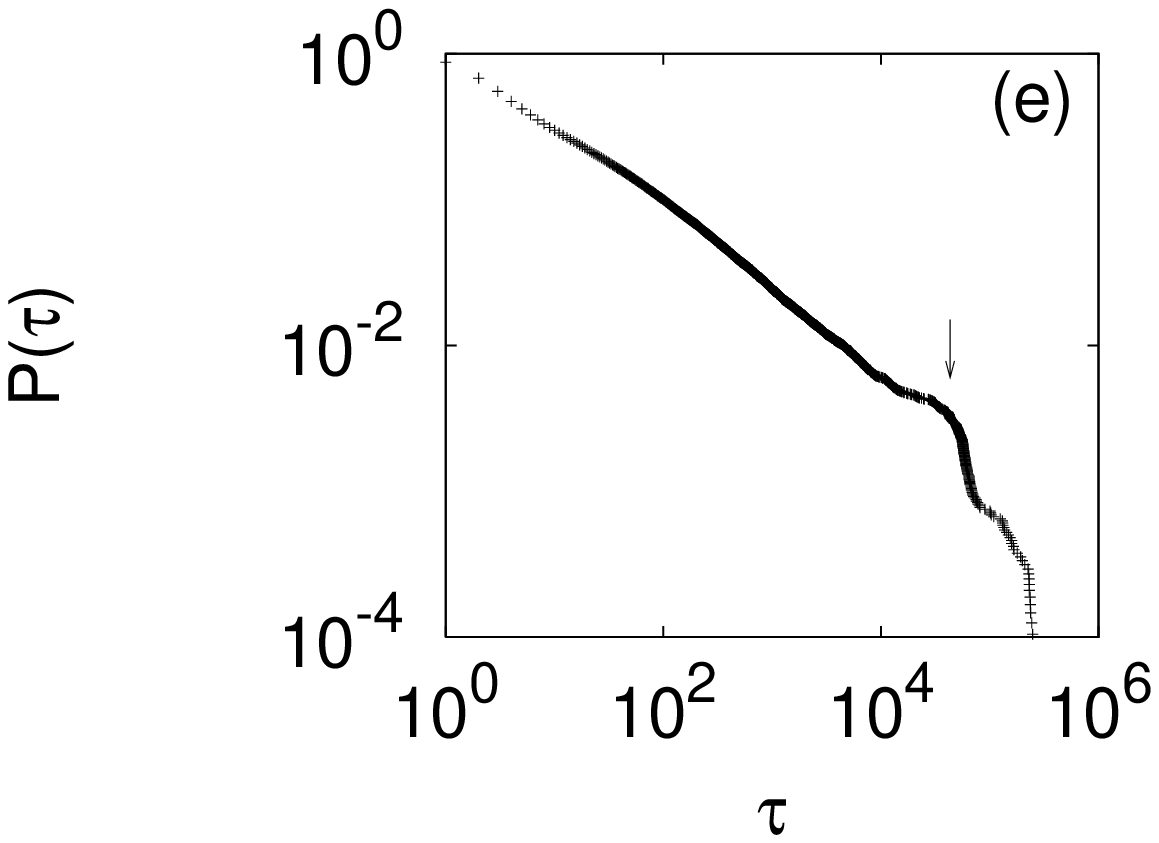}
\includegraphics[width=.5\textwidth]{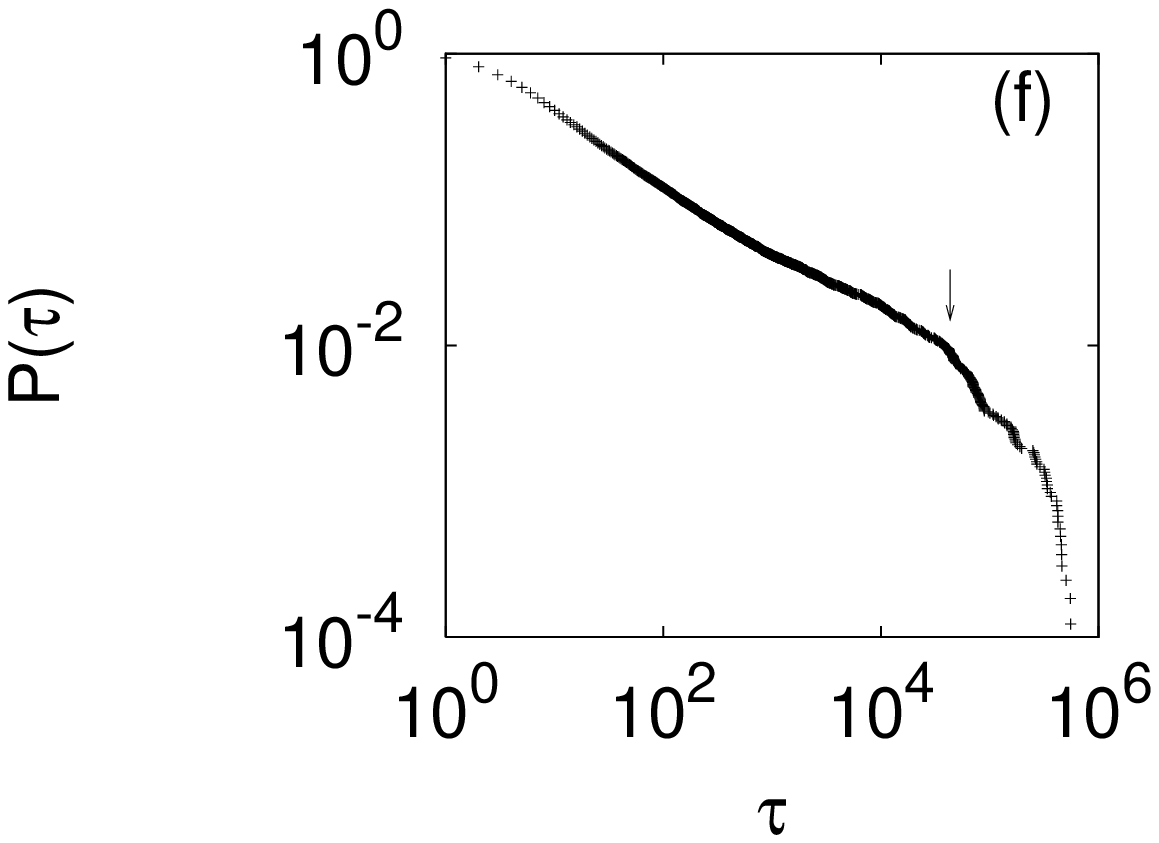}
\includegraphics[width=.5\textwidth]{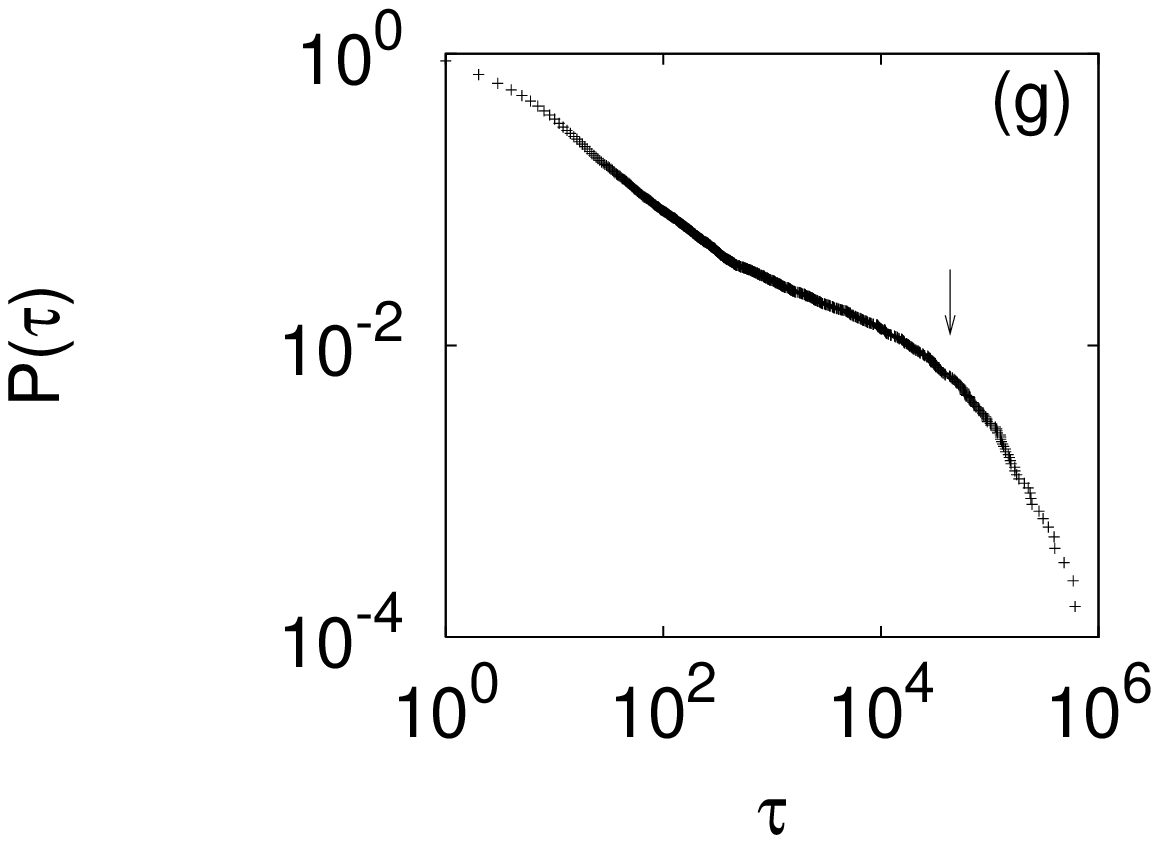}
\includegraphics[width=.5\textwidth]{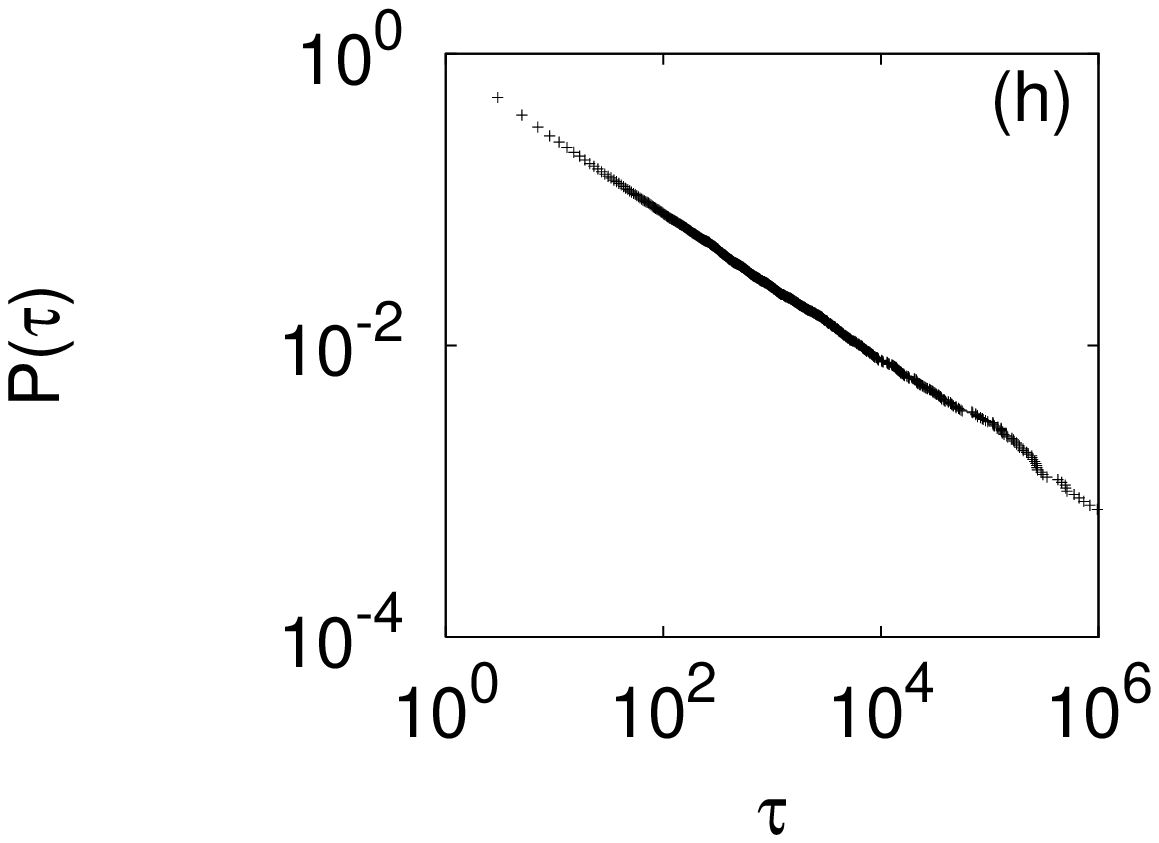}
\caption{Cumulative distributions of waiting times in collected datasets.
The panels from (a) to (g) correspond to the empirical datasets from $A$ to $G$,
respectively, while the panel (h) indicates the dataset $R$ from a random
walker.
Each arrow indicates $\tau=12$ h.}
\label{fig:cdf}
\end{figure}

Let us consider the probability distribution $p(\tau)$, obtained from an
empirical dataset, $\{\tau_i\}$. For convenience, we are going to
work with its derived form, the cumulative distribution function defined as
follows~\cite{newman}:
\[ P(\tau) = \int_\tau^\infty p(\tau')~d\tau'.\]
Fig.~\ref{fig:cdf} displays $P(\tau)$'s for our eight datasets.
The distribution function in  Fig.~\ref{fig:cdf}(d) looks far from a
straight line, presumably because this user prefers using his own personal
Linux machine to accessing the remote server where the recording has been
carried out.  In comparison, the distribution in Fig.~\ref{fig:cdf}(e)
which is for the same user as in (d) (he submitted two history files
$D$ from a remote server and $E$ from his own local desktop, respectively) 
does not show any significant difference from other datasets.
It evidently shows an effect of individual characteristics,
and the concave shape is reminiscent of the job submission interval
in supercomputers~\cite{kleban}.
Nevertheless, if we exclude the dataset $D$, 
the qualitative behaviors are surprisingly similar to each other.

\begin{figure}
\includegraphics[width=.8\textwidth]{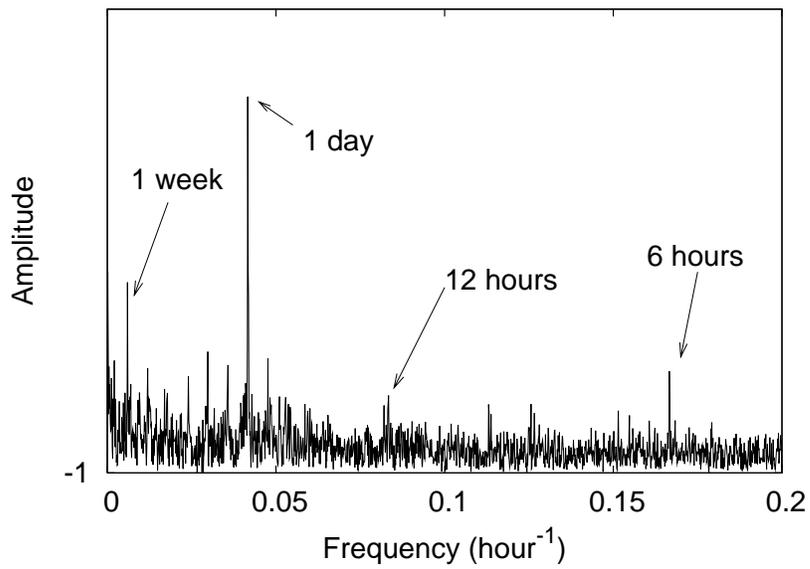}
\caption{Fourier transformation of input rates in the dataset $A$.}
\label{fig:fft}
\end{figure}

\begin{figure}
\includegraphics[width=.8\textwidth]{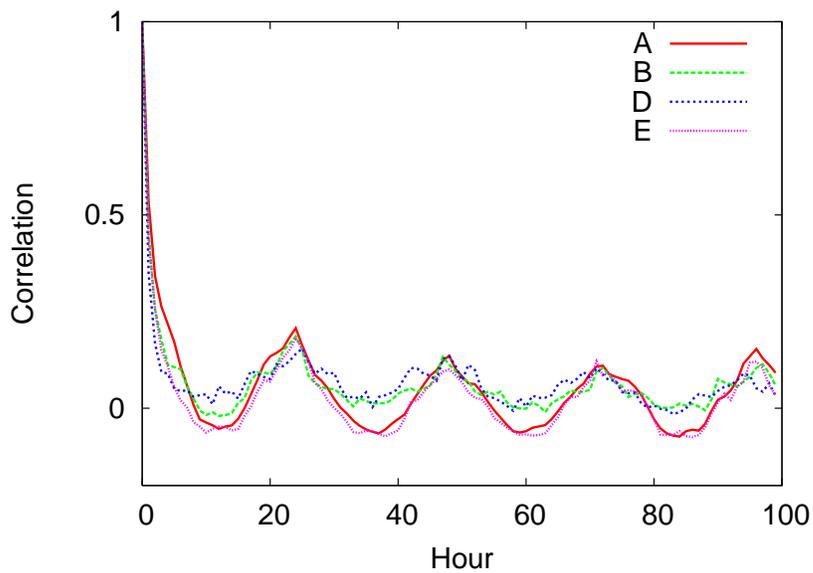}
\caption{(Color online) Autocorrelation of input rates in each dataset.
We depict only four datasets which clearly show oscillatory patterns with a
period of $24$ h.}
\label{fig:bcorr}
\end{figure}

Every arrow in Fig.~\ref{fig:cdf} indicates the point at
$4.32\times10^4$ s, or $12$ h. The existence
of a hump for each individual appears to reflect  his
daily life cycle. 
The Fourier transform also confirms that the working
pattern is quite regular, as two peaks are prominent at one day
and one week (Fig.~\ref{fig:fft}).
Shown differently, the autocorrelation function from the input
rate~\cite{kujawski} exhibits oscillatory patterns with periods
around $24$ h (Fig.~\ref{fig:bcorr}). All of these
indicate the existence of long-term orders.

We fit each dataset using a power-law function $p(\tau) \sim \tau^{-\alpha}$
with an appropriate lower bound $\tau_{\rm min}$, where the number of data
points larger than or equal to $\tau_{\rm min}$ is denoted as $n_{\rm tail}$.
The optimal parameter values are listed in Table~\ref{table:fit}.
We apply the goodness-of-fit test based on the Kolmogorov-Smirnov (KS)
statistic and measure the $p$-value, the probability
that a dataset was drawn from the hypothesized distribution~\cite{clauset}.
As shown by $p$-values  in Table~\ref{table:fit}, the power-law function is 
found to be at least a moderate description for all the datasets except $D$. 
The humps due to long-term orders do little harms in the test,
because the KS test tends to be sensitive to the deviations
in the body part, rather than those in the tail part with a much smaller
scale.
We do not treat the model selection problem~\cite{clauset} here, but
it would not be a big surprise if they converge to the L\'evy stable
distribution in the long run by the generalized Central Limit Theorem.

\subsection{Mutual information}

To further analyze behavioral patterns, some authors employ the conventional
autocorrelation function~\cite{harder}:
\[ a(j) = \frac{1}{n-j}\sum_{i=1}^{n-j}(\tau_i-\bar\tau)
(\tau_{i+j}-\bar\tau).\]
Note that this function assumes the well-behavedness of
statistical moments  such as
the average $\bar\tau$.
None of them are well defined for power-law
distributions with $\alpha \le 2$~\cite{newman}, and the large variation of
$\bar\tau$ in Table~\ref{table:basic} implies this deficiency.

We next try  to devise an alternative measure for the  correlation, which should
be zero for perfectly uncorrelated data.
A possible trick is reverting the
generation algorithm for power-law distributed random
numbers: If $r$ is a random number uniformly drawn from $[0,1)$,
the formula given as
\[ x = x_{\rm min} (1-r)^{-1/(\alpha-1)} \]
makes a power-law distribution $p(x)\sim x^{-\alpha}$ with a lower bound
$x_{\rm min}$~\cite{clauset}. Therefore, if $\{x_i\}$ is a set of 
power-law distributed random numbers, the inverse transformation
\[ r_i = 1-\left(\frac{x_i}{x_{\rm min}}\right)^{1 - \alpha} \]
will generate a set of random numbers uniformly distributed between
$[0,1)$.
In discrete case as ours, each $x_i$ is not mapped to a
unique point, but to a set of points ranged over
\[\left[ 1-\left( \frac{x_i -\frac{1}{2}}{x_{\rm min}-\frac{1}{2}}
\right)^{1-\alpha}, 1-\left( \frac{x_i+\frac{1}{2}}{x_{\rm min}-\frac{1}{2}}
\right)^{1-\alpha} \right).\]
Since every number within this range is equivalent, a reasonable
choice is to draw a point $r_i$ randomly within the interval.
This indeterminacy makes some fluctuations on the final
result, but this trick still works giving us consistent estimates.
The mutual information between consecutive points is then calculated
as~\cite{kantz}
\begin{equation}
I(r_{i+1};r_i) = \sum_{r_i} \sum_{r_{i+1}} p(r_{i+1},r_i) \log\left[
\frac{p(r_{i+1}, r_i)}{p(r_{i+1}) p(r_i)} \right],
\label{eq:mi}
\end{equation}
where $p(r_{i+1},r_i)$ means the joint probability density function of
$r_{i+1}$ and $r_i$.
Accordingly, if $r_i$ and $r_{i+1}$ are completely uncorrelated, 
$I$ takes the null value.

\begin{figure}
\includegraphics[width=.8\textwidth]{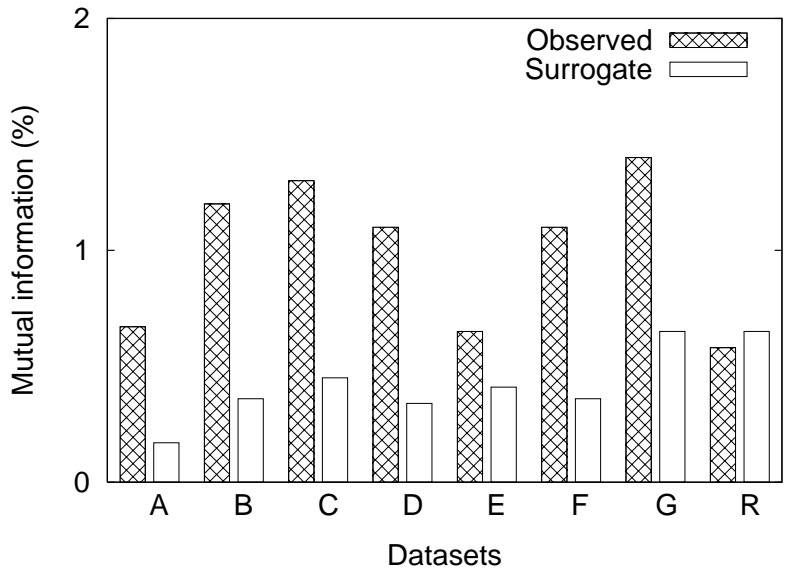}
\caption{Mutual information between consecutive time intervals for the
observed datasets and their surrogates.}
\label{fig:mi}
\end{figure}

Applying this transformation to each collected dataset, $\{\tau_i\}$, we
obtain the transformed result, $\{u_i\}$, whose number of elements is
$n_{\rm tail}$.
Introducing $H_1 = -\sum p(u_i) \log p(u_i)$ and $H_2 = -\sum
p(u_{i+1},u_i) \log p(u_{i+1},u_i)$
as the entropy of $\{u_i\}$ and the joint entropy of $\{u_{i+1},u_i\}$,
respectively, we rewrite Eq.~(\ref{eq:mi}) as follows:
\[ I(u_{i+1};u_i) = 2H_1 - H_2. \]
Normalizing this with respect to the entropy, we get the following quantity
to measure how much correlation a dataset contains:
\[ h = 1 - \frac{H_2}{2H_1}.\]
In practice, $p(u)~du$ is estimated by the number of data points between
$[u,u+du)$, and the values of entropies are dependent on the choice of
$du$, or equivalently, the number of bins in making a histogram.
We choose Sturges' formula to determine the optimal number of
bins~\cite{sturges}:
\[ k = \lceil \log_2 n_{\rm tail} + 1 \rceil, \]
where $\lceil x \rceil$ means the ceiling function of $x$.
The results are shown in Fig.~\ref{fig:mi}.
Every $h$ lies at around $1\%$, which is not much larger than our expectation.
However, we have to check if those values are small enough to conclude
that the data are indeed uncorrelated. A common technique to find a
reference point is by using
surrogates~\cite{kantz}: To destroy all the possible correlation without
altering the overall distribution, it suffices to perform a random
shuffle on the data.
Then we calculate the mutual information from an ensemble of such
surrogate datasets.
As shown in Fig.~\ref{fig:mi}, while this method makes little
differences in $R$, every other human dataset is found to carry mutual
information to a significant degree.
Therefore this implies that our datasets have differences from
what the previously proposed priority-based queue model predicts from the
viewpoint of mutual information.

Before proceeding, however, some subtlety should be
mentioned:
Since this test simply neglects all the $\tau_i$'s smaller
than $\tau_{\rm min}$, some pairs of $(u_i,u_{i+1})$ may not come from
the really consecutive time intervals. If we further require such
consecutiveness, the number of available data pairs becomes even less
than $n_{\rm tail}$, making the results also unclear for some datasets.
Therefore, even though we could reveal some quantitative differences,
they are not so conclusive as the constraint $\tau>\tau_{\rm min}$
and the indeterminacy for a discrete case severely worsen the power of
this test.
It is for the reason that we newly propose another test, taking all the
data points into consideration.

\subsection{Arcsine test}

\begin{figure}
\includegraphics[width=.8\textwidth]{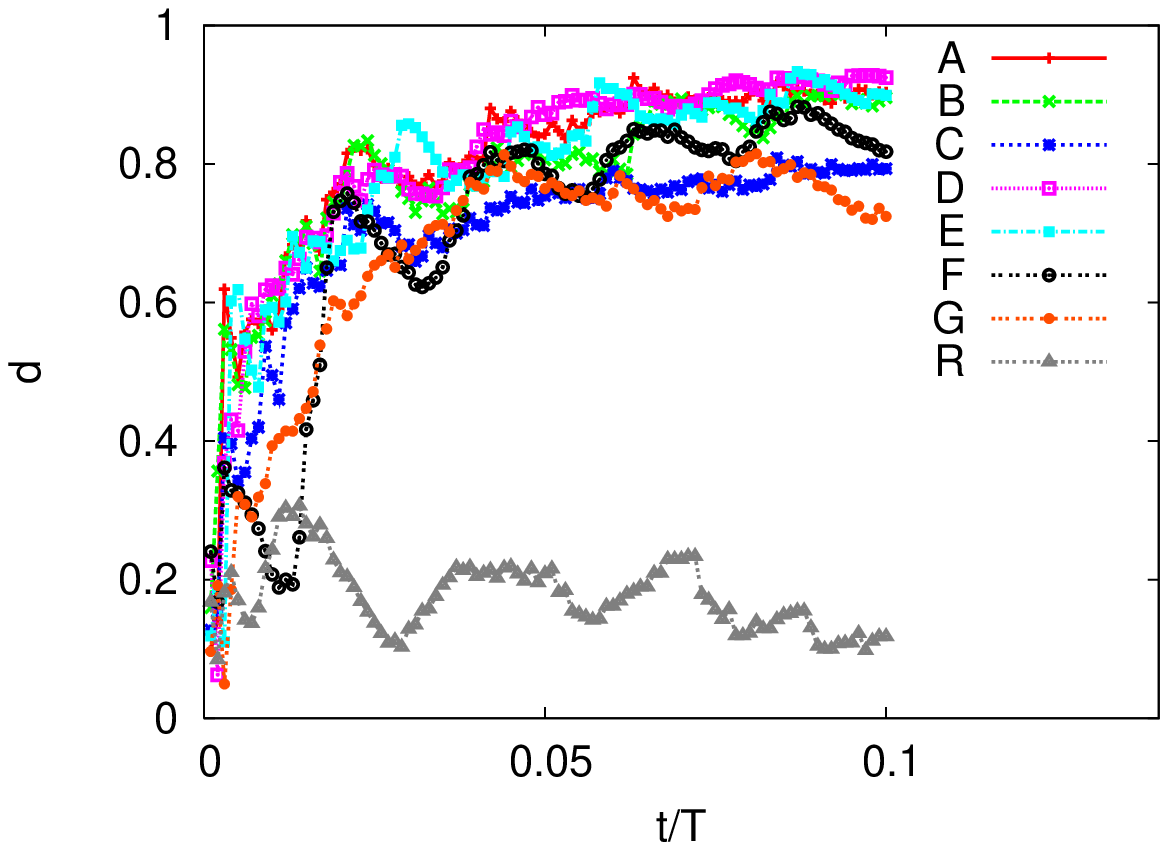}
\caption{(Color online) Deviation $d$ between the estimated probability function
and the arcsine law [see Eq.~(\ref{eq:d}) and text] as $t$ increases up to $10\%$ of 
the total recording period, $T$. Except for $R$ from the random walk, all
datasets are shown not to converge to the arcsine law.}
\label{fig:asin}
\end{figure}

Suppose that a one-dimensional random walker in the $x$-direction starts
at time $t=0$ from the origin at $x=0$. 
It is then mathematically proved that the probability $f$ for the walker
to hit the origin ($x=0$) in the time interval $(\xi t,t)$ with $0< \xi < 1$ 
is given by~\cite{ross} 
\begin{equation}
\label{eq:asin}
f(\xi) = 1-\frac{2}{\pi} {\rm arc}
\sin\sqrt{\xi} ,
\end{equation}
as $t \rightarrow \infty$.
Let us check how much our datasets are away from this result.
For a given $t$, we can estimate the probability function, $\eta =
f_e(\xi)$, from our datasets with varying $\xi$. It is a monotonically
decreasing function by definition, and the following KS statistic will
properly quantify the the maximum deviation~\cite{press}:
\begin{equation} 
\label{eq:d}
d = \max_\eta |f_e^{-1}(\eta) - f^{-1}(\eta)|. 
\end{equation} 
Note that since each measured point should contribute an equal amount in
the KS test, the inverse functions are more appropriate.
Due to the assumption that $t \rightarrow \infty$ which Eq.~(\ref{eq:asin}) 
is based on, we observe how the statistic $d$ behaves as $t$ increases
(Fig.~\ref{fig:asin}).
We only use the time $t$ less than $10\%$ of the total recording period
$T$ in order to avoid effects caused by  the finiteness of the time.
As clearly shown in Fig.~\ref{fig:asin}, only the dataset $R$ maintains
low $d$ at large $t$. Moreover, one may easily calculate the significance
level from the fact that $f_e(\xi)$ is constructed with the effective
number of points $N_e = 49$ (see Ref.~\cite{press} for details
of the KS test): The dataset $R$ has $d \approx 0.12$
at $t \approx 0.1T$ which amounts to the significance level of
$Q_{KS}\approx 46\%$.
Even if this does not satisfy the usual requirement like $95\%$
significance level, it is still remarkable since
we observe that every other human dataset has $Q_{KS}<10^{-20}$ under the
same condition.
Consequently, it is very plausible that the human dynamics 
in our datasets needs a modified description than a simple random walker,
which is also supported by the previous test based on the mutual information.

\section{Discussion}
\label{sec:discuss}

One interesting point in our observation is that the datasets show
heavy-tailed distributions up to some cutoffs and, at the same time,
quite regular behaviors.
This is seemingly contradictory, as a power-law distribution with an
exponent $\alpha \le 2$ is known to make its average and
variance diverge, lacking any characteristic time scales.

Even though one may say that the irregularity still exists in
intra-day scales at least (see Fig.~\ref{fig:cdf}),
it is true that a power-law distribution does not
necessarily imply highly complicated dynamics.
Let us consider a very simple example with
$N$ persons, each of whom has her own working frequency, $f_i~~
(i=1,\cdots,N)$. 
If these frequencies are uniformly distributed for a rather wide range of
time scales, the collected set of waiting times will have a power-law
distribution.
Namely, the number of each
person's own time interval is simply an inverse quantity of $f_i$,
and should have the following distribution~\cite{newman}:
\[ p(\tau) = p(f) \left| \frac{df}{d\tau} \right| = \frac{p(f)}{\tau^2}
\sim \tau^{-2}. \]
Even a single person may have multiple working phases, each of which
requires a different frequency of inputs but occupies roughly the same time
as other phases. Indeed, the exponent $\alpha\approx 2$ is already reported in
Refs.~\cite{tao,slijper}, and modulating the exponent is not impossible
because any random parts in fragmenting time schedules are basically
a multiplicative process yielding power-law or log-normal
distributions~\cite{sornette}.

This is an illustrative, if not serious, example to show that
there may be a number of competing theories, all of which yield
power-law distributions with being totally different in other
respects~\cite{newman}.
We have thus focused on consistency checks for a previously suggested
model, while leaving how to elaborate on a new one to be a future work.
We stress that rejecting the existing queue model in our case does not mean
that it is wrong or useless. Rather, our study shows one of its greatest
virtues, i.e., the openness to a variety of challenges.
Therefore, the queueing scheme is still a good starting point to consider
human behaviors at the first approximation in a variety of situations, once
one keeps in mind how a current simple model may deviate from reality.

\section{Conclusion}
\label{sec:conc}

We collected human behavioral patterns from Linux history files and found
that their waiting-time distributions followed power-laws.
Since they seemed to fall into the previously claimed universality class,
characterized by an exponent of $1.5$, we tested the corresponding
priority-based queue
model by two measures. The first test was based on the mutual information,
while the second was on the arcsine law in a regenerative process.
Both tests indicated that our datasets had significant differences from what
the model of our concern predicted.
This implies that we should also consider the temporal relations
in order to find an accurate description of human behaviors.

\begin{ack}
We are grateful to Y.-Y. Ahn, H. A. T. Kiet, S. H. Lee, and  J. Um for
providing us with their history files.
This work was supported by
the Korea Science and Engineering Foundation
through the Basic Research Program with grant No. R01-2007-000-20084-0.
\end{ack}


\begin{thebibliography}{10}
\expandafter\ifx\csname url\endcsname\relax
  \def\url#1{\texttt{#1}}\fi
\expandafter\ifx\csname urlprefix\endcsname\relax\def\urlprefix{URL }\fi

\bibitem{albert}
R.~Albert, H.~Jeong, A.-L. Barab\'asi, The diameter of the world wide web,
  Nature 401 (1999) 130.

\bibitem{kujawski}
B.~Kujawski, J.~Holyst, G.~J. Rodgers, Growing trees in Internet news groups
  and forums, Phys. Rev. E 76 (2007) 036103.

\bibitem{wang}
W.-X. Wang, B.-H. Wang, C.-Y. Yin, Y.-B. Xie, T.~Zhou, Traffic dynamics based
  on local routing protocol on a scale-free network, Phys. Rev. E 73 (2006)
  026111.

\bibitem{bjkim}
B.~J. Kim, S.~M. Park, Distribution of Korean family names, Physica A 347
  (2005) 683.

\bibitem{baek}
S.~K. Baek, H.~A.~T. Kiet, B.~J. Kim, Family name distributions: master
  equation approach, Phys. Rev. E 76 (2007) 046113.

\bibitem{kiet}
H.~A.~T. Kiet, S.~K. Baek, H.~Jeong, B.~J. Kim, Korean family name distribution
  in the past, J. Korean Phys. Soc. 51 (2007) 1812.


\bibitem{barabasi}
A.-L. Barab\'asi, The origin of bursts and heavy tails in human dynamics,
  Nature 435 (2005) 207.

\bibitem{oliveira}
J.~G. Oliveira, A.-L. Barab\'asi, Darwin and {E}instein correspondence
  patterns, Nature 437 (2005) 1251.

\bibitem{vazquez2005}
A.~V\'azquez, Exact results for the Barab\'asi model of human dynamics, Phys.
  Rev. Lett. 95 (2005) 248701.

\bibitem{vazquez2006}
A.~V\'azquez, J.~G. Oliveira, Z.~Dezs{\"o}, K.-I. Goh, I.~Kondor, A.-L.
  Barab\'asi, Modeling bursts and heavy tails in human dynamics, Phys. Rev. E
  73 (2006) 036127.

\bibitem{grinstein}
G.~Grinstein, R.~Linsker, Biased diffusion and universality in model queues,
  Phys. Rev. Lett. 97 (2006) 130201.

\bibitem{stouffer-barabasi-r}
D.~B. Stouffer, R.~D. Malmgren, L.~A.~N. Amaral, Comment on the origin of
  bursts and heavy tails in human dynamics, e-print arXiv: physics/0510216v1 (2005);
A.-L. Barab\'asi, K.-I. Goh, A.~V\'azquez, Reply to comment on ``the origin of
  bursts and heavy tails in human dynamics'', e-print arXiv: physics/0511186 (2005).
\bibitem{tao}
T.~Zhou, H.~Kiet, B. J. Kim, B.-H. Wang, Role of activity in human dynamics,
  e-print arXiv: 0711.4168 (2007).

\bibitem{Vazquez2006b}
A.~V\'azquez, Impact of memory on human dynamics, Physica A 373 (2006)
  747.

\bibitem{blanchard}
P.~Blanchard, M.-O. Hongler, Human activity modeling and Barab\'asi's queueing
  systems, e-print arXiv: cond-mat/0608156v1 (2006).

\bibitem{kleban}
S.~D. Kleban, S.~H. Clearwater, Hierarchical dynamics, interarrival times, and
  performance, Proc. SC2003 (2003).

\bibitem{slijper}
H.~P. Slijper, J.~M. Richter, J.~B.~J. Smeets, M.~A. Frens, The effects of
  pause software on the temporal characteristics of computer use, Ergonomics 50
  (2007) 178.

\bibitem{newman}
M.~E.~J. Newman, Power laws, {P}areto distributions and {Z}ipf's law, Contemp.
  Phys. 46 (2005) 323.

\bibitem{ross}
S.~M. Ross, Stochastic Processes, 2nd Edition, John Wiley \& Sons, New
  York, 1996.

\bibitem{clauset}
A.~Clauset, C.~R. Shalizi, M.~E.~J. Newman, Power-law distributions in
  empirical data, e-print arXiv:0706.1062 (2007).

\bibitem{harder}
U.~Harder, M.~Paczuski, Correlated dynamics in human printing behavior, Physica
  A 361 (2006) 329.

\bibitem{kantz}
H.~Kantz, T.~Schreiber, Nonlinear Time Series Analysis, Cambridge University
  Press, Cambridge, 1997.

\bibitem{sturges}
H.~A. Sturges, The choice of a class interval, J. Am. Stat. Assoc. 21 (1926)
  65.

\bibitem{press}
W.~H. Press, S.~A. Teukolsky, W.~T. Vetterling, B.~P. Flannery, Numerical
  Recipes in {C}, 2nd Edition, Cambridge University Press, New York, 2002.

\bibitem{sornette}
D.~Sornette, R.~Cont, Convergent multiplicative processes repelled from zero:
  Power laws and truncated power laws, J. Phys. I 7 (1997) 431.

\end{thebibliography}

\newpage

\begin{table}
\caption{Basic quantities:
Datasets were collected from seven history files (A--G) 
and R was generated by a simple
computer program simulating a one-dimensional random walk.
The columns titled as $n$ and $T$ are the number of recorded commands and
the total recording period, respectively. The third column gives the average
time interval $\bar\tau \equiv T/n$, while the last one shows the maximal
interval in each dataset.
}
\begin{tabular*}{\hsize}{@{\extracolsep{\fill}}ccccc}\hline
Dataset & $n$ & $T$ & $\bar\tau$ & $\tau_{\rm max}$\\\hline
A & $9.3\times10^4$ & $2.9\times10^7$ & $3.1\times10^2$ &
$1.1\times10^6$\\
B & $2.2\times10^4$ & $2.9\times10^7$ & $1.3\times10^3$ &
$1.0\times10^6$\\
C & $1.3\times10^4$ & $2.9\times10^7$ & $2.1\times10^3$ &
$2.4\times10^6$\\
D & $3.0\times10^4$ & $2.7\times10^7$ & $8.9\times10^2$ &
$1.4\times10^6$\\
E & $4.8\times10^4$ & $2.1\times10^7$ & $4.4\times10^2$ &
$5.2\times10^5$\\
F & $1.6\times10^4$ & $2.9\times10^7$ & $1.8\times10^3$ &
$1.8\times10^6$\\
G & $1.2\times10^4$ & $1.5\times10^7$ & $1.2\times10^3$ &
$1.2\times10^6$\\
R & $2.0\times10^4$ & $1.3\times10^8$ & $6.3\times10^3$ &
$3.0\times10^7$\\
\hline
\end{tabular*}
\label{table:basic}
\end{table}

\begin{table} 
\caption{Results of the goodness-of-fit test:
Each dataset is fitted to the power-law distribution with a lower bound
$\tau_{\rm min}$ and an exponent $\alpha$ by the KS test. The number of
points satisfying $\tau_i \ge \tau_{\rm min}$ is denoted as $n_{\rm tail}$,
and a $p$-value means the probability that the power-law hypothesis is
correct.
}
\begin{tabular*}{\hsize}{@{\extracolsep{\fill}}ccccc}\hline
Dataset & $\tau_{\rm min}$ & $\alpha$ & $n_{\rm tail}$ & $p$-value \\\hline
A &  60 & 1.74 & $1.1\times10^4$ & 0.47 \\
B &  24 & 1.47 & $4.4\times10^3$ & 0.34 \\
C &  32 & 1.50 & $3.1\times10^3$ & 0.61 \\
D &  13 & 1.57 & $4.8\times10^3$ & 0.00 \\
E & 174 & 1.62 & $3.5\times10^3$ & 0.62 \\
F &  25 & 1.48 & $3.9\times10^3$ & 0.22 \\
G &  26 & 1.52 & $2.2\times10^3$ & 0.20 \\
R &  36 & 1.50 & $1.9\times10^3$ & 0.84 \\
\hline
\end{tabular*}
\label{table:fit}
\end{table}

\end{document}